# Possible Orthorhombic Phase of $Ta_2O_5$ under High Pressures

Yan Gong[1,2], HuiMin Tang[3], Yong Yang[1,2,*], and Yoshiyuki Kawazoe[4,5]

1. Key Lab of Photovoltaic and Energy Conservation Materials, Institute of Solid State Physics, HFIPS, Chinese Academy of Sciences, Hefei 230031, China
2. Science Island Branch of Graduate School, University of Science and Technology of China, Hefei 230026, China
3. College of Physics and Technology, Guangxi Normal University, Guilin 541004, China
4. New Industry Creation Hatchery Center (NICHe), Tohoku University, 6-6-4 Aoba, Aramaki, Aoba-ku, Sendai, Miyagi 980-8579, Japan
5. Department of Physics and Nanotechnology, SRM Institute of Science and Technology, Kattankulathurm, 603203, TN, India

A potential orthorhombic phase of $Ta_2O_5$, designated as Y-$Ta_2O_5$, is predicted under high-pressure conditions through density functional theory (DFT) calculations combined with structural search algorithms. This phase, consisting of four formula units per unit cell (Z = 4), exhibits the highest known Ta-O coordination numbers. Y-$Ta_2O_5$ is found to be the most energetically favorable form of $Ta_2O_5$ in the pressure range of approximately 70 GPa to at least 200 GPa. Both standard DFT-GGA and higher-accuracy GW calculations reveal that Y-$Ta_2O_5$ is a wide bandgap semiconductor with a direct bandgap. Additionally, nuclear quantum effects (NQEs) introduce nontrivial corrections to external pressure at fixed volumes, underscoring their significance in high-pressure phase stability analyses.

*Corresponding Author: yyanglab@issp.ac.cn

Tantalum pentoxide ($Ta_2O_5$), a prototypical wide-bandgap semiconductor, is widely used in optoelectronic materials [1-5], resistive random-access memory (RRAM) [6-9], and catalysis [10-12] owing to its exceptional dielectric properties, high optical transparency, low reflectivity, and its intrinsic capability for electron-hole separations. These attributes render it particularly suitable for high-performance anti-reflection coatings that significantly enhance light absorption and photoelectric conversion efficiency in solar cells [13, 14]. Beyond conventional applications, $Ta_2O_5$ demonstrates remarkable potential as a UV photodetector material with applications in military sensing and related fields [15-17]. Furthermore, tantalum-based materials exhibit unique multifunctionality, combining superior biocompatibility, chemical inertness, antimicrobial properties, and excellent mechanical performance (including high hardness and wear resistance), making them strategically important for biomedical applications [18-22]. Recent breakthroughs in $Ta_2O_5$/multi-walled carbon nanotube (MWCNT) heterostructures have yielded photodetectors with substantially enhanced photocurrent and fast response speeds (rise and decay times < 80 ms). This exceptional performance highlights their potential for advanced applications in high-speed optical communications, precision optoelectronics, and real-time environmental monitoring [23].

Systematic experimental and theoretical studies have confirmed that the crystal structure of $Ta_2O_5$ exhibits significant environmental dependency. Substantial evidence suggests its phase composition is primarily regulated by temperature and pressure conditions, while preparation methods also significantly influence the final structure. Based on growth conditions, $Ta_2O_5$ crystal phases can be divided into two categories: (a) ambient-pressure phases and (b) high-pressure phases [24-26].

Regarding ambient-pressure phase research, significant progress has been made after years of investigation (Detailed in Supplementary Materials). The main ambient-pressure phases discovered so far include $L_{SR}$ (space group $Pm$, $Z = 11$, containing 22 Ta and 55 O atoms) [27], $L_G$ ($C112/m$, $Z = 19$) [28], $T$ ($Pmm2$, $Z = 12$) [29], $\delta$ ($P6/mmm$, $Z = 2$) [30], $\beta_{AL}$ ($Pccm$, $Z = 2$) [31], $\beta_R$ ($Z = 2$) [32], $\lambda$ ($Pbam$, $Z = 2$) [33], $\gamma$ ($P1$, $Z = 2$) [34], and $\gamma_1$ ($P\text{-}1$, $Z = 1$) [35].

In comparison, research on the high-pressure phases of $Ta_2O_5$ has progressed more slowly. Currently, only two high-pressure phases have been reported: B phase (*C2/c*, $Z = 4$) and Z phase (*C2*, $Z = 2$). The B phase and Z phase were successfully synthesized by Zibrov's team under conditions of 8 GPa and 1470 K, and characterized using X-ray powder diffraction and high-resolution transmission electron microscopy (HRTEM) [36]. Theoretical calculations by Pérez-Walton *et al.* confirmed that within the studied pressure range, the B phase exhibits the highest thermodynamic stability [24], but whether new phases exist at higher pressures remains unknown. On the other hand, the effects due to the quantum motions of atomic nuclei, or nuclear quantum effects (NQEs) [37-52], are generally neglected in previous studies under high-pressure conditions when the dynamics don't involve light elements (such as hydrogen and its isotopes).

This study employs first-principles calculations combined with structural search methods to successfully predict a possible new high-pressure phase of $Ta_2O_5$ (named as Y-$Ta_2O_5$, with a space group of *Ibam*, and $Z = 4$). Calculation results demonstrate that under pressures above 62 GPa, the Y phase shows superior thermodynamic stability compared to the B phase. Importantly, this study specifically considers the influence of nuclear quantum effects (NQEs) on external pressure, finding that one of the aspects of NQEs, i.e., zero-point energy (ZPE) leads to significant pressure corrections. This indicates that observable NQEs are not limited to the dynamical processes in systems containing light elements [37-52].

This work systematically investigated the crystal structures of the $Ta_2O_5$ system using first-principles calculations based on density functional theory (DFT). We performed a global structural search on the $Ta_2O_5$ crystal structures with four formula units (Z=4) by combining the VASP code [53, 54] with the USPEX crystal structure prediction program [55, 56] using an evolutionary algorithm (EA), while employing the VASP and CALYPSO code [57-60] with particle swarm optimization (PSO) to double-check the predicted results. This approach demonstrates significant advantages in identifying stable structures at specific stoichiometric ratios and has been widely employed in high-pressure research, providing crucial support for advancements in the field [61, 62]. In structural search, all the calculations adopted the PBE-type general gradient approximation (GGA) exchange-correlation functional [63] with a plane-wave cutoff energy of 600 eV and a k-point mesh spacing of 0.06 × 2π Å$^{-1}$. Through this approach, we successfully identified an orthorhombic structure with a space group of *Ibam*. To achieve higher accuracy of electronic structures, we have employed the GW method [64, 65] to correct the band gap values and verified the structural dynamical stability through phonon calculations based on density functional perturbation theory (DFPT) [66]. The phonon calculations used a 2×2×2 supercell with convergence criteria of 1×10$^{-9}$ eV for energy and 3×10$^{-4}$ eV/Å for forces. The computational details and additional information regarding the cross-validation between USPEX and CALYPSO results can be found in the Supplementary Materials.

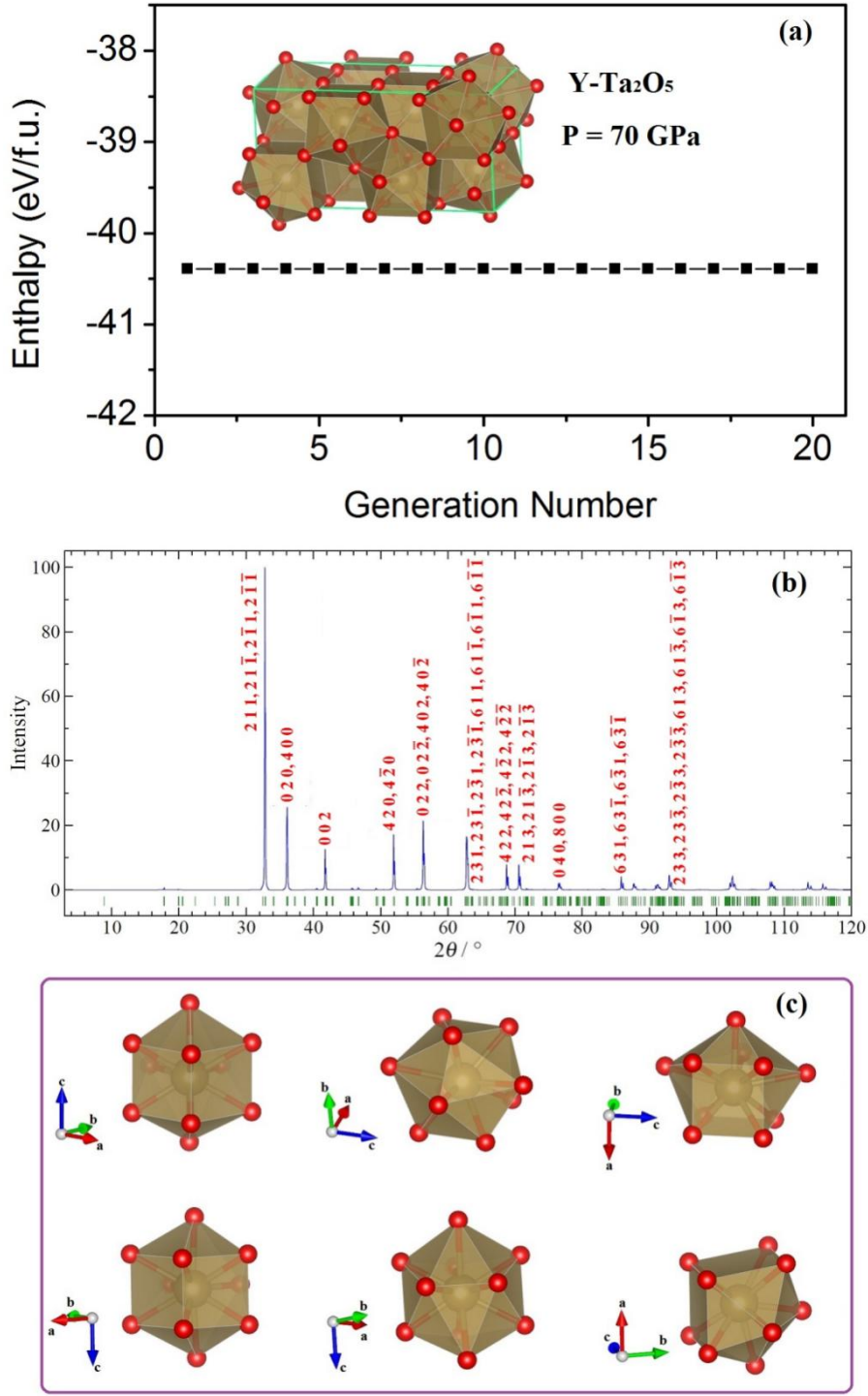

**Fig. 1.** (a) Enthalpy evolution with generation number for $Ta_2O_5$ ($Z = 4$) structures at P = 70 GPa, calculated using DFT in combination with evolutionary algorithm (EA); (b) Simulated X-ray diffraction pattern using Cu Kα1 radiation (λ = 1.541 Å); (c) Three-dimensional visualization of the fundamental $TaO_{10}$ polyhedral unit (brown spheres: Ta atoms; red spheres: O atoms).

Figure 1(a) presents the results of structural search for $Ta_2O_5$ at a pressure P = 70 GPa, showing excellent energy convergence after 20 generations of evolutionary optimization. The refined Y-phase $Ta_2O_5$ crystallizes in the orthorhombic system with an Ibam space group, having lattice parameters of $a$ = 9.939 Å, $b$ = 4.977 Å, $c$ = 4.322 Å ($\alpha = \beta = \gamma = 90.00°$). The corresponding atomic coordinates are presented in Supplementary Materials (Section 3.1). Equivalent structures are predicted by VASP + CALYPSO calculations (see Supplementary Materials, Section 3.2 & Table SI). The simulated X-ray diffraction pattern in Fig. 1(b) (Cu K$\alpha$1 radiation, $\lambda$ = 1.541 Å) displays its strongest peak at $2\theta$ = 32.81° ($d$-spacing = 2.728 Å) corresponding to (211), (21$\bar{1}$), (2$\bar{1}$1), and (2$\bar{1}\bar{1}$) planes, while the second and third most intense peaks appear at $2\theta$ = 36.06° ($d$-spacing = 2.488 Å, (020) and (400)) and 56.34° ($d$-spacing = 1.632 Å, (022), (02$\bar{2}$), (402), (40$\bar{2}$)), respectively. As can be seen from Fig. 1(c), compared with the B-phase [35, 36], the Y-phase exhibits higher Ta-O coordination numbers - the highest oxygen coordination number ($n$ = 10) known to date in metal oxides. Similar to B-$Ta_2O_5$, all the building blocks $TaO_n$ of Y-$Ta_2O_5$ are identical in geometry. Simulations using VASP + USPEX predict that the Y-phase maintain as the most stable structure to at least 200 GPa (see Supplementary Materials, Section 3.3 & Table SII for structural and energy parameters).

Based on enthalpy calculations (see Table 1 and detailed analysis below), we determined that the pressure for the structural transformation from the B-phase to the Y-phase is approximately 62 GPa. At $P$ = 62 GPa, although the zero-point energies (ZPE) of both phases are nearly identical (with the Y-phase slightly higher), and the internal energy of equilibrium lattice ($E_0$) of the B-phase is lower, the Y-phase has a smaller unit cell volume with more compact Ta-O stacking, which accounts for its lower enthalpy value.

**Table 1**. Unit cell and energy parameters describing the Y and B-$Ta_2O_5$ at external pressure $P$ = 62 GPa. The volume of unit cell ($V_{cell}$), the zero-point energies (ZPE), and the difference of internal energy and enthalpy with reference to Y-$Ta_2O_5$ ($\Delta E_0$, $\Delta H$) are listed.

| Ta$_2$O$_5$ | Y | B |
|---|---|---|
| a (Å) | 9.985 | 13.596 |
| b (Å) | 4.995 | 4.378 |
| c (Å) | 4.347 | 5.083 |
| α (°) | 90.00 | 90.00 |
| β (°) | 90.00 | 129.22 |
| γ (°) | 90.00 | 90.00 |
| Space group | Ibam (No.72) | C2/c (No.15) |
| TaO$_n$ | TaO$_{10}$ (×8) | TaO$_8$ (×8) |
| V$_{cell}$ (Å$^3$/f.u.) | 54.20 | 58.61 |
| ΔE$_0$ (eV/f.u.) | 0 | -1.69 |
| ZPE (eV/f.u.) | 0.66 | 0.65 |
| ΔH (eV/f.u.) | 0 | 0.01 |

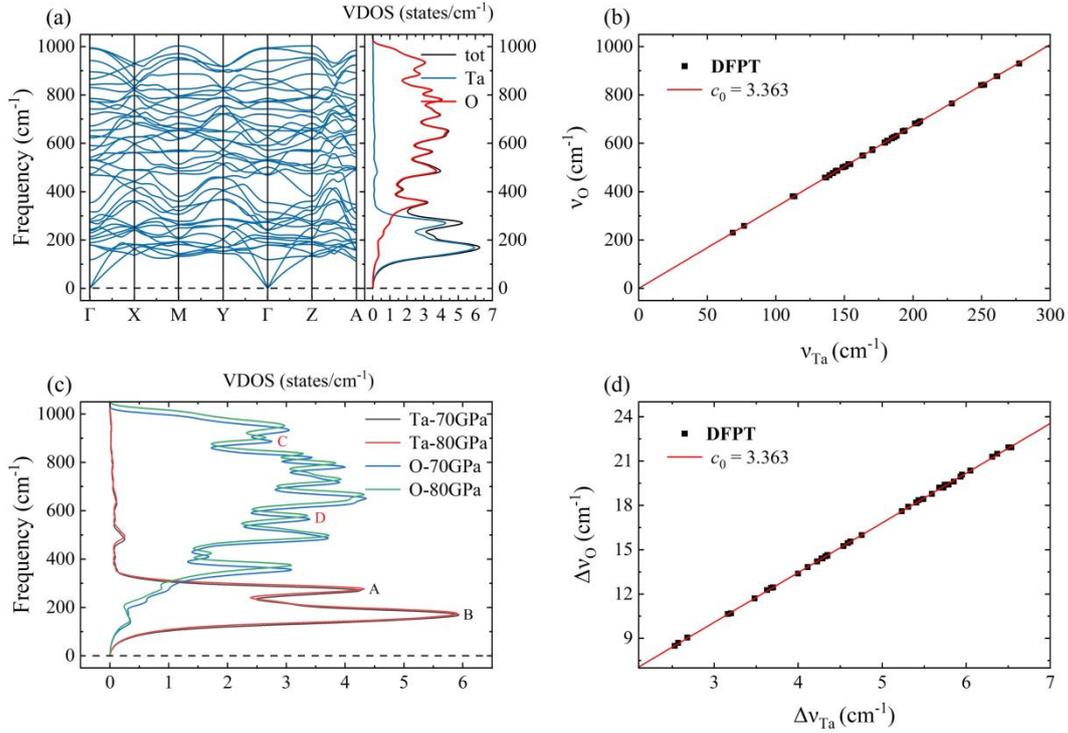

**Fig. 2.** (a) Calculated phonon dispersion relations (left panel) along high-symmetry q-point paths in the Brillouin zone (BZ) and vibrational density of states (VDOS, right panel) for Y-Ta$_2$O$_5$ at 70 GPa. The direct coordinates of q-points in the BZ are defined as: Γ = (0, 0, 0), X = (0.5, 0, 0), M = (0.5, 0.5, 0), Y = (0, 0.5, 0), Z = (0, 0, 0.5), A = (0.5, 0.5, 0.5); (b) Frequency correspondence of Ta- and O-dominated vibrational modes at 70 GPa; (c) Atom-resolved VDOS and pressure-induced blueshifts at $P$ = 70,

80 GPa; (d) The frequency blueshift correspondence of Ta- and O-dominated vibrational modes under different pressures.

To study the dynamical stability and lattice vibration characteristics of Y-$Ta_2O_5$, we conducted systematic phonon spectrum calculations, based on density functional perturbation theory (DFPT). This approach enables accurate calculation of the system's phonon properties directly at the primitive cell scale, significantly reducing computational costs [67]. Figure 2(a) displays the phonon dispersion curves along high-symmetry directions in the Brillouin zone and the corresponding vibrational density of states (VDOS) of Y-$Ta_2O_5$. Notably, no imaginary frequencies are observed throughout the Brillouin zone for either acoustic or optical branches, strongly confirming the excellent dynamic stability of the Y-$Ta_2O_5$ structure.

Through atom-projected VDOS analysis [Figs. 2(b)-(c)], we observed distinct frequency partitioning in the vibrational modes of Ta and O atoms: the heavier Ta atoms (relative atomic mass 180.95 amu) predominantly contribute to low-frequency vibrational modes ($v < 300$ cm$^{-1}$), while the lighter O atoms (relative atomic mass 16.00 amu) dominate the high-frequency region ($v > 300$ cm$^{-1}$). This observation can be explained within the model of harmonic oscillator. Generally, for crystals whose primitive cells contain different types of atoms, when the vibrations of one type of atom dominate while the vibrations of other types can be neglected, it can be shown under the harmonic approximation that the vibration frequency $v$ is inversely proportional to the square root of the atomic mass $M$ [68]: $v \propto \frac{1}{\sqrt{M}}$. For $Ta_2O_5$, the frequency of each Ta-dominated optical mode and that of the O-dominated optical modes can be correlated one-to-one. Shown in Fig. 2(b), are the frequencies of a number of Ta-dominated optical modes and O-dominated optical modes, where excellent agreement with the prediction of harmonic approximation is found. For example, the two VDOS peaks (A, B) marked in Fig. 2(c) which mainly due to the vibrations of Ta atoms, find their correspondence of two peaks (C, D) in the high frequency region. Details of the vibrational modes are provided in the Supplementary

Materials (Fig. S1). Furthermore, increased pressure leads to decreased Ta-O distances and therefore stronger interatomic interactions (i.e., force constants) and blueshifts of frequencies, as demonstrated by Fig. 2(c) where the external pressure increases from 70 GPa to 80 GPa. It can be shown that the magnitude of frequency variation ($\Delta\nu$) is also inversely proportional to the square root of the atomic mass $M$ ($\Delta\nu \propto \frac{1}{\sqrt{M}}$). Based on the mass ratio of Ta (180.95 amu) to O (16.00 amu), the theoretically predicted frequency ratio $c_0 = \nu_O/\nu_{Ta} = (M_O/M_{Ta})^{-1/2} \approx 3.363$, the lines with this slope correspond well with the DFPT frequencies and frequency blueshifts data shown in Figs. 2(b) and 2(d). This explains why heavier Ta atoms tend to exhibit low-frequency vibrations, while lighter O atoms dominate the high-frequency modes.

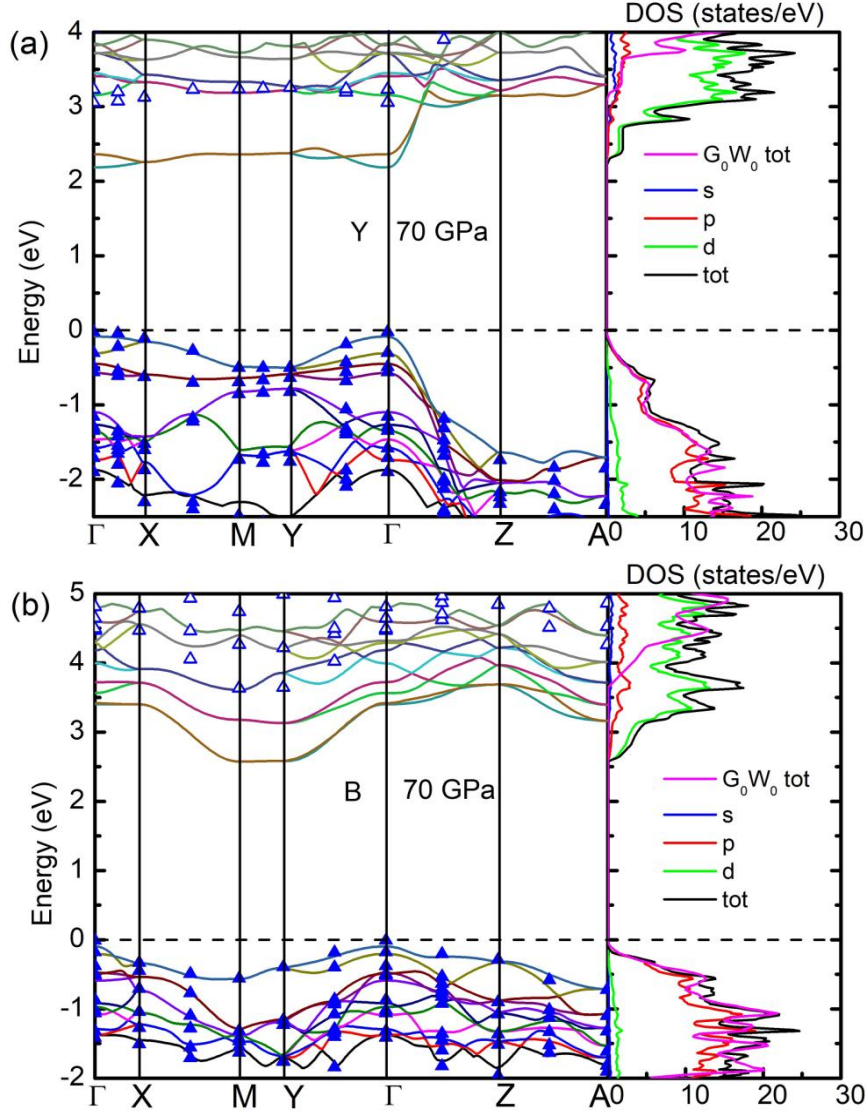

Fig. 3. The electron band structures, density of states (DOS), and partial density of states (PDOS) for Y (upper panels) and B-$Ta_2O_5$ (lower panels) at $P = 70$ GPa, calculated using both DFT-GGA and GW method. In the band structure plot, the solid lines represent the GGA energy bands along high-symmetry k-point paths, while the scattered solid and hollow blue triangles denote the GW results. The Fermi level (indicated by the vertical dashed line) is set at 0 eV.

**Table 2.** The valence band maximum (VBM) and conduction band minimum (CBM) and band gap of the two $Ta_2O_5$ phases (Y, B) at $P = 70$ GPa, obtained using DFT-GW method. The DFT-GGA data are listed in parentheses.

| Ta$_2$O$_5$ | VBM (eV) | CBM (eV) | Eg (eV) |
|---|---|---|---|
| Y | 9.29 (8.81) | 12.46 (11.08) | 3.17 (2.27) |
| B | 7.17 (6.78) | 10.81 (9.45) | 3.64 (2.67) |

Figure 3 presents a comprehensive comparison of the electronic structures of Y-Ta$_2$O$_5$ and B-Ta$_2$O$_5$ under high-pressure conditions. Through combined DFT-GGA (PBE) and GW calculations, we identify distinct electronic behaviors: the Y-phase exhibits a direct bandgap at Γ point (GW: 3.17 eV; PBE: 2.27 eV), contrasting with the B-phase's indirect bandgap characteristic with Γ→M transitions (GW: 3.64 eV; PBE: 2.67 eV). Remarkably, both the valence band maximum (VBM = 9.29 eV) and conduction band minimum (CBM = 12.46 eV) in the Y-phase show significant upward shifts of +2.12 eV and +1.65 eV respectively relative to the B-phase values (VBM = 7.17 eV; CBM = 10.81 eV). These electronic structure modifications originate from pronounced pressure-enhanced orbital hybridization and electron cloud redistribution effects in the Y-phase. While the Y-phase displays a marginally smaller bandgap (by 0.47 eV) compared to the B-phase, its elevated energy level alignment suggests enhanced charge transport capabilities under high pressure. These findings provide crucial insights for the possibility of developing advanced high-pressure electronic devices.

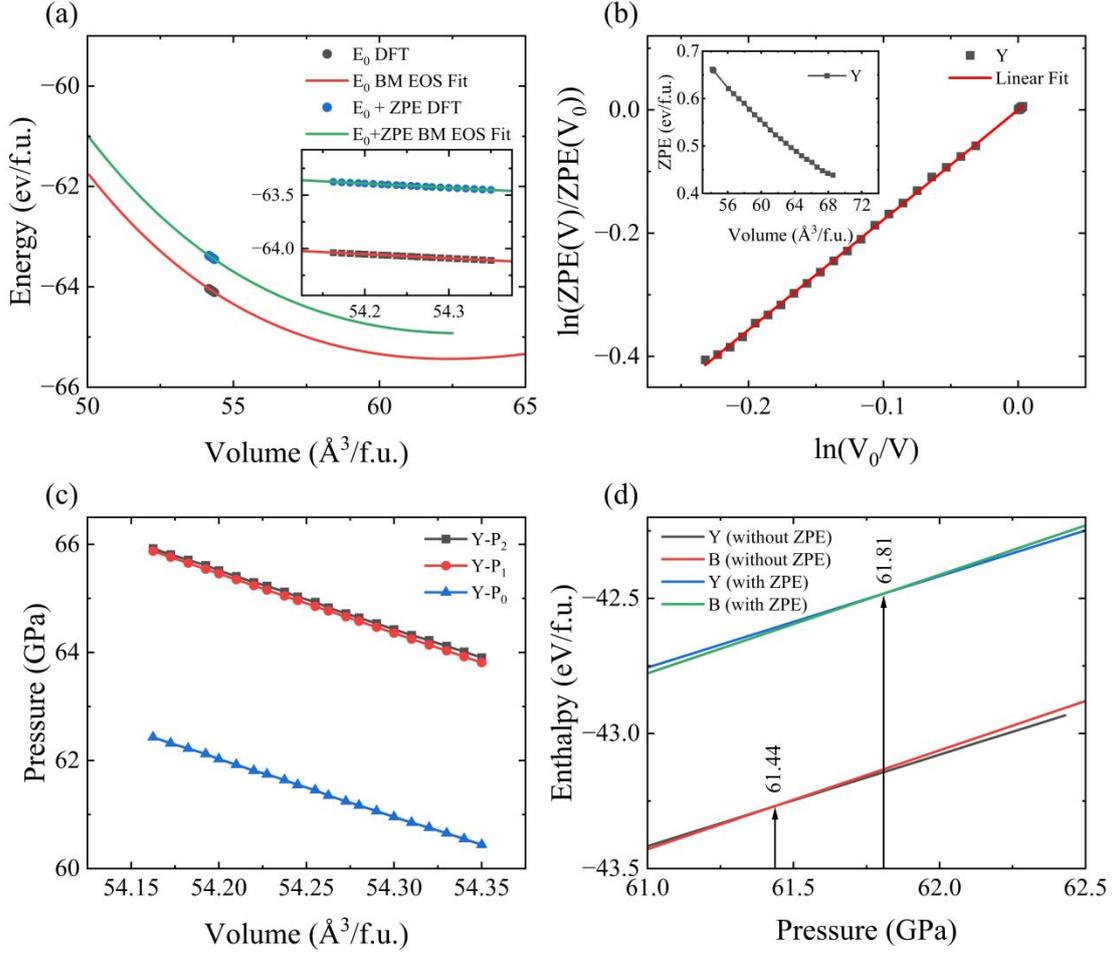

Fig. 4. (a) Total energy versus volume for Y-$Ta_2O_5$ near the phase transition point. The discrete points represent data calculated using DFT at different volumes. These data were fitted according to the third-order Birch-Murnaghan equation of state; (b) The dependence of ZPE with volume; (c) External pressure and its ZPE correction as functions of volume, where the subscripts 1 and 2 denote the data derived using the first and second scheme, respectively; (d) Enthalpies of the Y and B phase as a function of pressure, with the critical pressure point indicated, shown with and without ZPE correction.

The quantization of collective vibrations of atoms (more precisely, atomic nuclei) in crystal lattices manifests itself in the form of phonons [69]. The notable blueshifts of vibrational frequencies displayed in Fig. 2 imply the possibility of studying NQEs under high pressures. Generally, the external pressure exerted on a given crystal with volume $V$ at temperature $T$ may be calculated as follows [70]:

$$P = -\frac{dU}{dV} - \frac{dE_{ph}}{dV} = -\frac{dU}{dV} - \sum_j \left(\frac{1}{2}\hbar + \frac{\hbar}{e^{\hbar\omega_j/(k_BT)}-1}\right)\frac{d\omega_j}{dV} = P_0 + \Delta P \qquad (1)$$

where $U$ is the internal energy of equilibrium lattice at volume $V$, which can be readily obtained using DFT calculations, and $P_0 = -\frac{dU}{dV}$ is the pressure without taking into account NQEs; the phonon energy $E_{ph} = \sum_j \left(\frac{1}{2}\hbar\omega_j + \frac{\hbar\omega_j}{e^{\hbar\omega_j/(k_BT)}-1}\right)$, $\hbar$ is the reduced Planck constant and $\omega_j$ is the frequency of the $j$th mode. The correction to pressure due to NQEs is then given by

$$\Delta P = -\frac{dE_{ph}}{dV} = -\sum_j \left(\frac{1}{2}\hbar + \frac{\hbar}{e^{\hbar\omega_j/(k_BT)}-1}\right)\frac{d\omega_j}{dV}$$

$$= -\sum_j \left(\frac{1}{2}\hbar\omega_j + \frac{\hbar\omega_j}{e^{\hbar\omega_j/(k_BT)}-1}\right)\frac{1}{V} \times \frac{dln\omega_j}{dlnV} \qquad (2)$$

In room-temperature and below, the phonon energy is predominantly contributed by ZPE, i.e., $E_{ph} \cong \sum_j \frac{1}{2}\hbar\omega_j = ZPE$. Consequently, the correction to pressure due to lattice vibrations can be expressed as follows:

$$P \cong -\frac{dU}{dV} - \frac{dZPE}{dV} = P_0 + \Delta P$$

$$= P_0 - \frac{1}{2}\hbar\sum_j \left(\frac{d\omega_j}{dV}\right) = P_0 - \sum_j \left(\frac{1}{2}\hbar\omega_j\right)\frac{1}{V} \times \frac{dln\omega_j}{dlnV} \qquad (3)$$

To evaluate the role of NQEs, we proceed to calculate the pressures at target volumes via two numerical schemes. In the first scheme, we calculated the total energy $E_{tot}$ of Y-Ta$_2$O$_5$ at different volumes $V$, and the $E_{tot} \sim V$ data are least-squares fitted to the third-order Birch-Murnaghan (BM) equation of state (EOS) [71, 72] using VASPKIT [73]:

$$E_{tot}(V) = E_{tot0} + \frac{9V_0B_0}{16}\left\{\left[\left(\frac{V_0}{V}\right)^{2/3}-1\right]^3 B_0' + \left[\left(\frac{V_0}{V}\right)^{2/3}-1\right]^2 \left[6-4\left(\frac{V_0}{V}\right)^{2/3}\right]\right\}, \qquad (4)$$

where the terms $E_{tot0}$ and $V_0$ represent the total energy and volume at equilibrium (the external pressure $P = 0$), respectively; and $B_0$ and $B_0'$ represent the bulk modulus and its pressure derivative at equilibrium volume $V_0$ at which $E_{tot}(V)$ reaches its minimum. The role of NQEs is studied when the term $E_{tot0}$ takes into account the ZPE of at each volume. The parameters $B_0$ and $B_0'$ for the third-order BM EOS of the Y phase are listed in Supplementary Materials (Table SIII). Compared with the first-order Murnaghan EOS [74], the third-order BM EOS significantly improves fitting accuracy in high-pressure regimes by incorporating nonlinear terms, establishing itself as a widely used formula for studying material behavior under high pressures [75]. The pressure is determined using the third-order BM EOS:

$$P = -\left(\frac{\partial E_{tot}}{\partial V}\right) = \frac{3B_0}{2}\left[\left(\frac{V_0}{V}\right)^{7/3} - \left(\frac{V_0}{V}\right)^{5/3}\right] \times \left\{1 + \frac{3}{4}(B_0' - 4)\left[\left(\frac{V_0}{V}\right)^{2/3} - 1\right]\right\} \quad (5)$$

In the second scheme, we utilize the approximation by assuming that for all modes $\omega_j$ the term $-\frac{d\ln\omega_j}{d\ln V} \cong \gamma$, which is the Grüneisen parameter. Then Eq. (1) can be rewritten as follows:

$$P \cong -\frac{dU}{dV} + \frac{\gamma}{V}\sum_j \left(\frac{1}{2}\hbar\omega_j + \frac{\hbar\omega_j}{e^{\hbar\omega_j/(k_B T)} - 1}\right) = -\frac{dU}{dV} + \frac{\gamma}{V}E_{ph} = P_0 + \Delta P, \quad (6)$$

where $P_0$ is readily obtained by DFT calculations and the second term ($\Delta P$) can be deduced using the Grüneisen parameter and phonon energy. In the room-temperature and cryogenic regimes, $E_{ph} \cong ZPE$, and therefore one has $\Delta P \cong \frac{\gamma}{V}ZPE$.

We go further to demonstrate the pressure correction due to NQEs around the critical point of phase transition from B to Y phase. The total energies with and without ZPE corrections versus volume are shown in Fig. 4(a), where good agreement is found for the DFT data and their numerical fitting curves, validating the applicability of third-order BM EOS across this pressure regime. Figure 4(b) plots the dependence of ln(ZPE(V)/ZPE(V$_0$)) with ln(V$_0$/V), where pronounced linear relation is found, and the slope of which yields a Grüneisen parameter to be γ ~ 1.78. An approximately linear drop of ZPE with volume is also found in the in the vicinity of the phase transition (inset of Fig. 4(b)). The pressures computed using the two schemes (labeled as $P_1$ and $P_2$, respectively) to include NQEs are shown in Fig. 4(c), together with the uncorrected data ($P_0$). The data in Fig. 4(c) reveal remarkable consistency between the two independent approaches for determining ZPE-induced pressure corrections: EOS fitting and Grüneisen approximation, with the correction magnitude remaining essentially constant (~ 3.5 GPa). From Eq. (3), $P \cong -\frac{dE_0}{dV} - \frac{dZPE}{dV} = P_0 + \Delta P$, if the ZPE follows an approximately linear relationship with volume

variation — as illustrated in the inset of Fig. 4(b) with $\frac{dZPE}{dV} = k_0$ — then $\Delta P = -k_0$, remains constant. This accounts for the nearly invariant differences between $P_0$ and $P_1$, $P_2$ at varying volumes, as shown in Fig. 4(c).

We next examine the influence of ZPE on the critical pressure of phase transition. The enthalpy of each phase is given by $H = (E_0 + ZPE) + PV$. Phase transition occurs when the enthalpies of two competing phases equals. The pressure at which this condition is met defines the critical pressure $P_c$ (See Supplementary Materials):

$$P_c = -\frac{\Delta E_0}{\Delta V} - \frac{\Delta ZPE}{\Delta V} = P_{c0} + \Delta P_c \qquad (7)$$

Thus, the pressure correction $\Delta P_c = -\frac{\Delta ZPE}{\Delta V}$ induced by ZPE primarily originates from the differences of ZPE and volume at the transition point. As the ZPE difference increases and volume difference decreases, significant enhancement in the pressure correction $\Delta P_c$ due to NQEs can be expected. In the vicinity of B-to-Y transition under high pressures, the enthalpy data in Fig. 4(d) show that $P_{c0} \cong 61.44$ GPa, $P_c \cong 61.81$ GPa, and $\Delta P_c = 0.37$ GPa. From Table 1, the volume changes by $\Delta V \cong -4.41$ Å$^3$/f.u., and the ZPE changes by $\Delta ZPE \cong 0.01$ eV/f.u., which gives the critical pressure $\Delta P_c = -\frac{\Delta ZPE}{\Delta V} \cong 0.36$ GPa. The agreement is excellent.

To summarize, we predicted a high-pressure phase of Ta$_2$O$_5$ (Y-Ta$_2$O$_5$) with four formula units per crystal unit cell (Z=4) using first-principles calculations combined with structural search algorithms. Our calculations show that B-Ta$_2$O$_5$ undergoes a transition to this new high-pressure polymorph at ~ 62 GPa, and Y-Ta$_2$O$_5$ remains as thermodynamically the most stable phase up to at least 200 GPa. Electronic structure calculations reveal that Y-Ta$_2$O$_5$ exhibits direct bandgap semiconducting behavior. We have further investigated the mutual relationship between external pressure and phonon vibrations. Near the critical phase transition pressure, our calculations demonstrate that zero-point energy (ZPE) corrections to the pressure values are non-negligible. This finding provides clear evidence that observable nuclear quantum effects (NQEs) exist even in non-light-element systems like transition metal oxides under high-pressure conditions.

## Acknowledgements

This work is financially supported by the National Natural Science Foundation of China (No. 12074382, 11474285, 12464012). We would like to thank Professor Enge Wang for reading and helpful comments on the manuscript. We are grateful to the staff of the Hefei Branch of Supercomputing Center of Chinese Academy of Sciences, and the Hefei Advanced Computing Center for support of supercomputing facilities. We also thank the staff of the Center for Computational Materials Science, Institute for Materials Research, Tohoku University, and the supercomputer resources through the HPCI System Research Project (hp200246).

# Supporting Materials for "Possible Orthorhombic Phase of Ta$_2$O$_5$ under High Pressures"

Yan Gong[1,2], HuiMin Tang[3], Yong Yang[1,2*], Yoshiyuki Kawazoe[4,5]

*1. Key Lab of Photovoltaic and Energy Conservation Materials, Institute of Solid State Physics, HFIPS, Chinese Academy of Sciences, Hefei 230031, China*

*2. Science Island Branch of Graduate School, University of Science and Technology of China, Hefei 230026, China*

*3. College of Physics and Technology, Guangxi Normal University, Guilin 541004, China*

*4. New Industry Creation Hatchery Center (NICHe), Tohoku University, 6-6-4 Aoba, Aramaki, Aoba-ku, Sendai, Miyagi 980-8579, Japan*

*5. Department of Physics and Nanotechnology, SRM Institute of Science and Technology, Kattankulathurm, 603203, TN, India*

\*Email: yyanglab@issp.ac.cn

## 1. A Brief Review on the Studies of Ambient-pressure Phases of Ta$_2$O$_5$:

The $L$ phase has two structures: $L_{SR}$ and $L_G$. The $L_{SR}$ phase is an orthorhombic system structure ($Pm$, $Z=11$, containing 22 Ta and 55 O atoms) proposed by Stephenson and Roth [1]; the $L_G$ phase is a single crystal synthesized by Grey et al. using the flux method ($C112/m$, $Z=19$) [2].

The $T$ phase was established by Hummel et al. based on a modified Stephenson-Roth model, belonging to the orthorhombic system (Pmm2, $Z=12$) [3].

The $\delta$ phase is a hexagonal system structure ($P6/mmm$, $Z=2$) predicted by Fukumoto and Miwa through first-principles calculations [4].

The $\beta$ phase has three models: $\beta_{AL}$ orthorhombic phase ($Pccm$, $Z=2$) determined by Aleshina and Loginova using Rietveld full-pattern refinement [5]; simplified orthorhombic $\beta_R$ phase proposed by Ramprasad [6]; and the high-symmetry orthorhombic phase $\lambda$ ($Pbam$, $Z=2$) developed by Lee et al [7].

Notably, Yang and Kawazoe employed evolutionary algorithms combined with first-principles calculations to discover the significantly more stable $\gamma$ triclinic phase ($P1$, $Z=2$) [8], which was subsequently verified by Tong et al. who further predicted the existence of $\gamma_1$ triclinic phase ($P$-$1$, $Z=1$) using particle swarm optimization algorithms [9].

## 2. Details of Theoretical Methods

### 2.1 Structural Search and Optimization

In our *ab initio* simulations using VASP and USPEX package, we constructed $Ta_2O_5$ systems using the following evolutionary strategy: the initial generation consisted of 160 randomly generated symmetry-constrained structures, while the subsequent 19 generations were produced through operations including genetic crossover, atomic substitution, and structural mutation, with each generation maintaining ~ 50 candidate structures.

To ensure the reliability of our predictions, we independently verified the structures obtained from EA using VASP combined with the CALYPSO code. Specifically, we similarly constructed a $Ta_2O_5$ system with 4 formula units ($Z = 4$), generating 50 candidate structures per generation. All structures in the first generation were randomly generated. In subsequent generations, 30 structures were produced via particle swarm optimization, while the remaining 20 structures were randomly generated under symmetry constraints.

During the structural search process, all the optimization calculations for crystal structures were performed using VASP. The calculations employed plane-wave basis sets to expand the electronic wave functions, with the projector augmented wave (PAW) method [10, 11] accurately treating the interactions between ion cores and valence electrons. To ensure computational accuracy, we set the plane-wave cutoff energy to 600 eV and adopted Monkhorst-Pack k-mesh [12] for Brillouin zone integration, with the grid spacing controlled at a density of $0.06 \times 2\pi$ Å$^{-1}$. The exchange-correlation energy was treated using the generalized gradient approximation (GGA) in the Perdew-Burke-Ernzerhof (PBE) [13] form.

The optimized results demonstrate that this lowest-energy structure exhibits orthorhombic characteristics, belonging to the Ibam space group (No. 72 in the International Tables for Crystallography). These refined structural parameters establish a reliable foundation for subsequent electronic structure and physical property calculations.

In this study, we conducted structural optimization and energy calculations on the reported high-pressure B-Ta$_2$O$_5$ phase to validate its stability, employing consistent accuracy standards as those used for the newly discovered Y-phase. Both phases were subjected to precise structural optimization using a 2×4×4 Monkhorst-Pack k-point mesh configuration.

### 2.2 Phonon Calculations

Through systematic calculations based on density functional perturbation theory (DFPT) [14], we performed comprehensive investigations into the dynamical stability of the obtained structures. Prior to phonon calculations, meticulous structural optimization was performed to guarantee total energy convergence to within $1\times10^{-9}$ eV per unit cell and atomic forces below 0.03 eV/Å.

A 2×2×2 supercell was constructed for dynamical matrix calculations, with the PHONOPY package [15] employed to fully solve the Hessian matrix and obtain both phonon dispersion relations and vibrational density of states. This rigorous phonon calculation protocol provides robust theoretical validation for assessing the material's dynamical stability, while the maintained consistency in computational parameters throughout ensures the reliability of comparative analyses between different polymorphs.

### 2.3 GW calculations

Considering that standard DFT-GGA methods generally suffer from bandgap underestimation, this study further employs the advanced GW approximation [16, 17] for quasiparticle energy level calculations. By accurately treating electron many-body interactions (including exchange-correlation effects), this method yields more precise band structures, demonstrating particular advantages for bandgap prediction in semiconductors and insulating materials.

For the specific computations, we implemented the one-shot GW method (G$_0$W$_0$) within the VASP package [18] to obtain the energy spectra. The quasiparticle energies were determined by solving the following equation [16]:

$$(T + V_{ext} + V_H)\psi_{nk}(\vec{r}) + \int d\vec{r}^{'} \sum (\vec{r},\vec{r}^{'};E_{nk}) \psi_{nk}(\vec{r}^{'}) = E_{nk}\psi_{nk}(\vec{r}) \qquad (S1)$$

In this expression, $T$ denotes the kinetic energy operator of electrons, $V_{ext}$

represents the external potential induced by ions, $V_H$ stands for the electrostatic Hartree potential, and $\Sigma$ is the self-energy operator. $E_{nk}$ and $\psi_{nk}(\vec{r})$ correspond to the quasiparticle energy and wavefunction, respectively. Within the GW approximation [17], the self-energy $\Sigma$ is computed as follows:

$$\Sigma(\vec{r},\vec{r}';E) = \frac{i}{2\pi}\int d\omega e^{i\delta\omega} G(\vec{r},\vec{r}';E+\omega)W(\vec{r},\vec{r}';\omega) \quad \text{(S2)}$$

where $G$ represents the Green's function, $W$ denotes the dynamically screened Coulomb interaction, and $\delta$ is an infinitesimal positive constant.

In the GW calculations, the plane-wave energy cutoff was set to 600 eV. For GGA calculations, k-point meshes of 2×4×4 for both Y-$Ta_2O_5$ and B-$Ta_2O_5$ were employed to ensure numerical convergence. For GW calculations, consistent k-point grids (4×8×8 for Y-$Ta_2O_5$ and 2×4×4 for B-$Ta_2O_5$) were implemented, with 320 and 256 energy bands included, respectively.

This parameter configuration guarantees that the calculated bandgap errors remain within ~ 0.1 eV (see Fig. S2 for convergence tests of Y and B-phase). This multi-scale computational strategy—spanning from structural optimization to electronic property calculations—ensures the reliability and accuracy of our findings throughout the entire research process.

## 3. The Structural Search Results of Y-$Ta_2O_5$

### 3.1 POSCAR of Y-$Ta_2O_5$ at P = 70 GPa, predicted using VASP+USPEX

```
Y-Ta2O5 P= 70 GPa VASP+USPEX
   1.00000000000000
     9.9389832239755940    0.0000000000000000    0.0000000000000000
     0.0000000000000000    4.9769584023627358    0.0000000000000000
     0.0000000000000000    0.0000000000000000    4.3219174610041184
   Ta   O
    8    20
Direct
     0.8791837709893745    0.7556414271571685    0.0000000000000000
     0.1208162140106278    0.2443585438428326    0.0000000000000000
     0.6208162290106255    0.2556414561571674    0.0000000000000000
```

|   |   |   |
|---|---|---|
| 0.3791837709893745 | 0.7443585728428315 | 0.0000000000000000 |
| 0.3791837709893745 | 0.2556414561571674 | 0.5000000000000000 |
| 0.6208162290106255 | 0.7443585728428315 | 0.5000000000000000 |
| 0.1208162140106278 | 0.7556414271571685 | 0.5000000000000000 |
| 0.8791837709893745 | 0.2443585438428326 | 0.5000000000000000 |
| 0.4194995632606084 | 0.3498251497024832 | 0.0000000000000000 |
| 0.5805004367393916 | 0.6501748502975168 | 0.0000000000000000 |
| 0.0805004367393916 | 0.8498251497024832 | 0.0000000000000000 |
| 0.9194995632606084 | 0.1501748502975168 | 0.0000000000000000 |
| 0.9194995632606084 | 0.8498251497024832 | 0.5000000000000000 |
| 0.0805004367393916 | 0.1501748502975168 | 0.5000000000000000 |
| 0.5805004367393916 | 0.3498251497024832 | 0.5000000000000000 |
| 0.4194995632606084 | 0.6501748502975168 | 0.5000000000000000 |
| 0.7346961244157543 | 0.0000000000000000 | 0.2500000000000000 |
| 0.2653038465842397 | 0.0000000000000000 | 0.7500000000000000 |
| 0.2653038465842397 | 0.0000000000000000 | 0.2500000000000000 |
| 0.7346961244157543 | 0.0000000000000000 | 0.7500000000000000 |
| 0.7653038755842457 | 0.5000000000000000 | 0.7500000000000000 |
| 0.2346961534157603 | 0.5000000000000000 | 0.2500000000000000 |
| 0.2346961534157603 | 0.5000000000000000 | 0.7500000000000000 |
| 0.7653038755842457 | 0.5000000000000000 | 0.2500000000000000 |
| 0.5000000000000000 | 0.0000000000000000 | 0.2500000000000000 |
| 0.5000000000000000 | 0.0000000000000000 | 0.7500000000000000 |
| 0.0000000000000000 | 0.5000000000000000 | 0.7500000000000000 |
| 0.0000000000000000 | 0.5000000000000000 | 0.2500000000000000 |

### 3.2 POSCAR of Y-Ta$_2$O$_5$ at P = 70 GPa, predicted using VASP+CALYPSO

Y-Ta$_2$O$_5$ P= 70 GPa VASP+CALYPSO
 1.0
         4.9766659        0.0000000        0.0000000
         0.0000000        9.9397133        0.0000000
         0.0000000        0.0000000        4.3224782
  Ta   O
  8   20
Direct
         0.74454          0.37937          0.00000
         0.74454          0.62063          0.50000
         0.25546          0.62063          0.00000
         0.25546          0.37937          0.50000
         0.24454          0.87937          0.50000
         0.24454          0.12063          0.00000
         0.75546          0.12063          0.50000

| | | |
|---|---|---|
| 0.75546 | 0.87937 | 0.00000 |
| 0.00000 | 0.73499 | 0.75000 |
| 0.35002 | 0.41972 | 0.00000 |
| 0.64998 | 0.58028 | 0.00000 |
| 0.50000 | 0.76501 | 0.75000 |
| 0.50000 | 0.76501 | 0.25000 |
| 0.64998 | 0.41972 | 0.50000 |
| 0.00000 | 0.50000 | 0.75000 |
| 0.00000 | 0.73499 | 0.25000 |
| 0.00000 | 0.50000 | 0.25000 |
| 0.35002 | 0.58028 | 0.50000 |
| 0.50000 | 0.23499 | 0.25000 |
| 0.85002 | 0.91972 | 0.50000 |
| 0.14998 | 0.08028 | 0.50000 |
| 0.00000 | 0.26501 | 0.25000 |
| 0.00000 | 0.26501 | 0.75000 |
| 0.14998 | 0.91972 | 0.00000 |
| 0.50000 | 0.00000 | 0.25000 |
| 0.50000 | 0.23499 | 0.75000 |
| 0.50000 | 0.00000 | 0.75000 |
| 0.85002 | 0.08028 | 0.00000 |

**Table SI Structure search using VASP + CALYPSO to identify the first twenty lowest-energy structures.**

| No. | space group | Enthalpy (eV/f.u.) | Volume (Å³/f.u.) |
|---|---|---|---|
| 1 | Ibam (No.72) | -40.3926 | 53.4546 |
| 2 | Ibam (No.72) | -40.3851 | 53.6644 |
| 3 | Ibam (No.72) | -40.3714 | 53.4729 |
| 4 | P-1 (No.2) | -40.3209 | 55.7380 |
| 5 | P-1 (No.2) | -40.3135 | 55.6257 |
| 6 | P1 (No.1) | -40.2959 | 55.6962 |
| 7 | P-1 (No.2) | -40.2945 | 55.7206 |
| 8 | P1 (No.1) | -40.2942 | 55.7206 |
| 9 | P1 (No.1) | -40.2847 | 55.7889 |
| 10 | C2/c (No.15) | -40.2309 | 55.3153 |
| 11 | P-1 (No.2) | -40.2229 | 54.7767 |
| 12 | P1 (No.1) | -40.2091 | 55.3933 |
| 13 | P-1 (No.2) | -40.1994 | 55.1171 |
| 14 | P1 (No.1) | -40.1876 | 55.3642 |
| 15 | P1 (No.1) | -40.1765 | 55.0220 |
| 16 | P-1 (No.2) | -40.1635 | 55.3254 |
| 17 | P-1 (No.2) | -40.1431 | 55.3807 |
| 18 | P-1 (No.2) | -40.1300 | 55.9454 |
| 19 | P1 (No.1) | -40.1298 | 55.6722 |
| 20 | P-1 (No.2) | -40.1294 | 55.4546 |

Table SI presents the first 20 lowest-energy configurations predicted by VASP+CALYPSO. It is noteworthy that the obtained ground-state structure shows good agreement with results obtained using VASP and USPEX.

### 3.3 POSCAR of Y-$Ta_2O_5$ at P = 200 GPa, predicted using VASP+USPEX

```
Y-Ta2O5 P=200 GPa VASP+USPEX
   1.0
        9.4334001541        0.0000000000        0.0000000000
        0.0000000000        4.7699999809        0.0000000000
        0.0000000000        0.0000000000        4.0749998093
   Ta   O
   8    20
Direct
        0.621600032         0.257900000         0.000000000
        0.378399998         0.742100000         0.000000000
        0.878399968         0.757900000         0.000000000
        0.121600002         0.242100000         0.000000000
        0.121600002         0.757900000         0.500000000
        0.878399968         0.242100000         0.500000000
        0.378399998         0.257900000         0.500000000
        0.621600032         0.742100000         0.500000000
        0.500000000         0.000000000         0.250000000
        0.500000000         0.000000000         0.750000000
        0.000000000         0.500000000         0.750000000
        0.000000000         0.500000000         0.250000000
        0.582499981         0.655300021         0.000000000
        0.417499989         0.344700009         0.000000000
        0.917500019         0.155299991         0.000000000
        0.082500011         0.844699979         0.000000000
        0.082500011         0.155299991         0.500000000
```

| | | |
|---|---|---|
| 0.917500019 | 0.844699979 | 0.500000000 |
| 0.417499989 | 0.655300021 | 0.500000000 |
| 0.582499981 | 0.344700009 | 0.500000000 |
| 0.735300004 | 0.000000000 | 0.250000000 |
| 0.264699996 | 0.000000000 | 0.750000000 |
| 0.264699996 | 0.000000000 | 0.250000000 |
| 0.735300004 | 0.000000000 | 0.750000000 |
| 0.764699996 | 0.500000000 | 0.750000000 |
| 0.235300004 | 0.500000000 | 0.250000000 |
| 0.235300004 | 0.500000000 | 0.750000000 |
| 0.764699996 | 0.500000000 | 0.250000000 |

**Table SII Structure search calculations using VASP + USPEX at 200 GPa**

| Generation | space group | Enthalpy (eV/f.u.) | Volume (Å³/f.u.) | Density (g/cm³) |
|---|---|---|---|---|
| 1 | Ibam (No.72) | -0.4540 | 45.8450 | 16.0060 |
| 2 | Ibam (No.72) | -0.4540 | 45.8450 | 16.0060 |
| 3 | Ibam (No.72) | -0.4540 | 45.8450 | 16.0060 |
| 4 | Ibam (No.72) | -0.4540 | 45.8380 | 16.0080 |
| 5 | Ibam (No.72) | -0.4540 | 45.8380 | 16.0080 |
| 6 | Ibam (No.72) | -0.4540 | 45.8380 | 16.0080 |
| 7 | Ibam (No.72) | -0.4540 | 45.8380 | 16.0080 |
| 8 | Ibam (No.72) | -0.4540 | 45.8380 | 16.0080 |
| 9 | Ibam (No.72) | -0.4540 | 45.8380 | 16.0080 |
| 10 | Ibam (No.72) | -0.4540 | 45.8360 | 16.0090 |
| 11 | Ibam (No.72) | -0.4540 | 45.8360 | 16.0090 |
| 12 | Ibam (No.72) | -0.4540 | 45.8360 | 16.0090 |
| 13 | Ibam (No.72) | -0.4540 | 45.8360 | 16.0090 |
| 14 | Ibam (No.72) | -0.4540 | 45.8360 | 16.0090 |
| 15 | Ibam (No.72) | -0.4540 | 45.8360 | 16.0090 |
| 16 | Ibam (No.72) | -0.4540 | 45.8360 | 16.0090 |

Our structure search study conducted at 200 GPa using the VASP + USPEX package reveals that the Y-phase remains stable under this pressure condition, exhibiting a negative enthalpy ($H < 0$).

## 4. Eigenvector Vector Diagrams of Typical Vibration Modes

Through post-processing analysis using Phonopy coupled with VESTA, we obtained the eigenvectors of optical vibration modes for the Y-phase under 70 GPa pressure. Representative peaks A, B, C and D were labeled, with their eigenvector vector diagrams of vibration modes presented in Fig. S1.

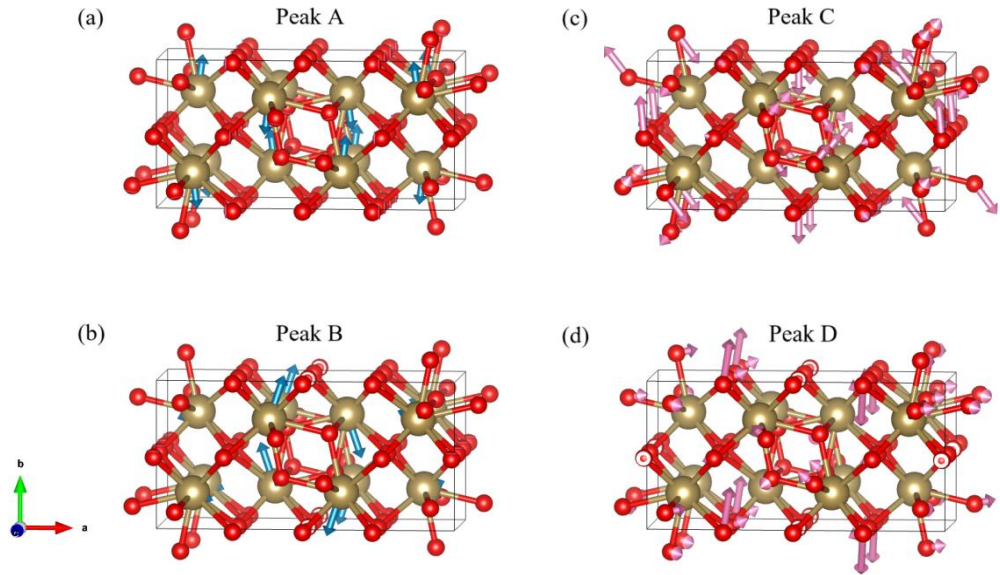

**Fig. S1.** (a) Peak A (70 GPa, $\nu = 277.03\ cm^{-1}$): Dominant eigenvector view of the Ta-dominated optical vibrational mode; (b) Peak B (70 GPa, $\nu = 172.40\ cm^{-1}$): Dominant eigenvector view of the Ta-dominated optical vibrational mode; (c) Peak C (70 GPa, $\nu = 929.99\ cm^{-1}$): Dominant eigenvector view of the O-dominated optical vibrational mode; (d) Peak D (70 GPa, $\nu = 579.25\ cm^{-1}$): Dominant eigenvector view of the O-dominated optical vibrational mode.

## 5. Table SIII Parameters $B_0$ and $B_0^{'}$ of the third-order Birch–Murnaghan equation of state (EOS) for the Y-phase

| $Ta_2O_5$ | $B_0$ (GPa) | $B_0^{'}$ |
|---|---|---|
| $Y_1$ | 312.7948 | 4.9077 |
| Y2 | 309.3143 | 4.7765 |

$Y_1$ is obtained from fitting the $E_{tot0} - V$ curve without $ZPE$, while $Y_2$ is derived from the $ZPE$-corrected $(E_{tot0} + ZPE) - V$ curve fitting.

## 6. The Derivation of Critical Pressure with ZPE Correction

Let subscripts 1 and 2 denote the two phases. At a given pressure $P$, the enthalpy values are $H_1 = (E_{01} + ZPE_1) + PV_1$ and $H_2 = (E_{02} + ZPE_2) + PV_2$. When the phase transition occurs, $H_1 = H_2$, yielding

$$(E_{01} + ZPE_1) + PV_1 = (E_{02} + ZPE_2) + PV_2. \quad (S3)$$

The critical pressure $P_c$ is then defined:

$$P_c(V_2 - V_1) = (E_{01} - E_{02}) + (ZPE_1 - ZPE_2) \quad (S4)$$

Let $\Delta V = (V_2 - V_1)$, $\Delta E_0 = (E_{02} - E_{01})$, $\Delta ZPE = (ZPE_2 - ZPE_1)$, Then:

$$P_c = -\frac{\Delta E_0}{\Delta V} - \frac{\Delta ZPE}{\Delta V} = P_{c0} + \Delta P_c \quad (S5)$$

## 7. Convergence Tests on the Key Parameters of GW Calculations

We provide convergence test on the key parameters (the number of energy bands, k-meshes) employed in our one-shot $G_0W_0$ calculation for Y-$Ta_2O_5$.

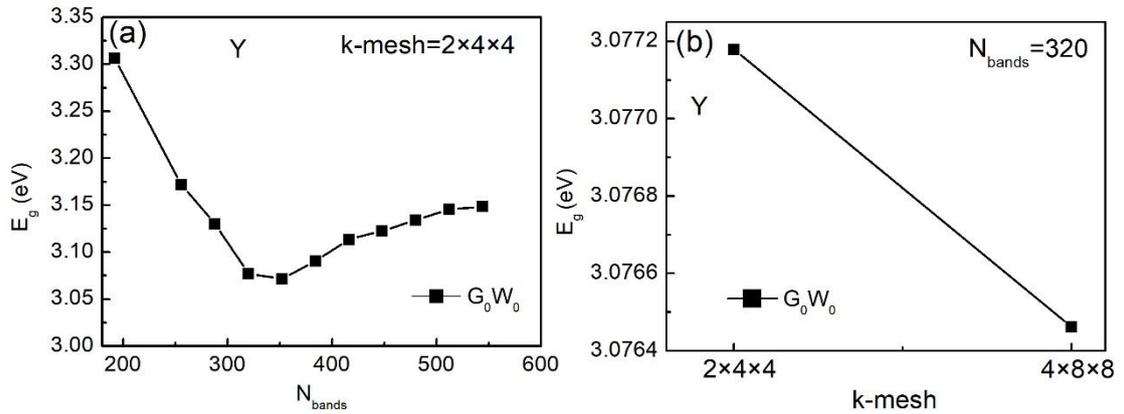

**Fig. S2.** The band gap ($E_g$) of Y-$Ta_2O_5$ as a function of the total number of energy bands (a) and k-meshes (b) used in the one-shot $G_0W_0$ calculation.